\documentclass[a4paper,11pt]{article}
\usepackage{epsfig}
\usepackage{graphicx}

\makeatletter

\@addtoreset{equation}{section}
\makeatother
\newcommand{\eq}{\begin{equation}}
\newcommand{\eqx}{\end{equation}}
\newcommand{\eqs}{\begin{equation*}}
\newcommand{\eqsx}{\end{equation*}}
\newcommand{\eqn}{\begin{eqnarray}}
\newcommand{\eqnx}{\end{eqnarray}}
\newcommand{\alg}{\begin{align}}
\newcommand{\algx}{\end{align}}

\newcommand{\f}[2]{\frac{#1}{#2}}

\newcommand{\lm}{\lambda}

\newcommand{\al}{\alpha}

\newcommand{\om}{\omega}
\newcommand{\gm}{\gamma}

\newcommand{\qqqq}{\quad\quad\quad\quad}
\newcommand{\st}{\tilde{s}}

\begin{document}
\begin{titlepage}
\vskip1cm
\begin{flushright}
UOSTP {\tt 110801}
\end{flushright}
\vskip1.25cm
\centerline{\Large
\bf Janus Black Holes }
\vskip1.25cm \centerline{  Dongsu Bak$^a$, Michael Gutperle$^b$ and  Romuald A. Janik$^c$ }
\vspace{1.25cm} \centerline{${}^a$  Physics Department,
University of Seoul, { Seoul 130-743, Korea}} \vskip0.15cm
\centerline{${}^b$  Department of Physics and Astronomy}
\centerline{\sl
University of California, Los Angeles, CA 90095,
USA} \vskip0.15cm \centerline{${}^c$  Institute of Physics, Jagiellonian University }
\centerline{\sl Reymonta 4, 30-059 Krak\'ow, Poland} \vskip0.25cm \centerline{\tt dsbak@uos.ac.kr,
gutperle@ucla.edu, romuald@th.if.uj.edu.pl } \vspace{1.5cm}
\centerline{\bf Abstract} \vspace{0.75cm} \noindent

In this paper Janus black holes  in $AdS_3$ are considered.
These are static solutions of an Einstein-scalar system with broken
translation symmetry along the horizon.
These solutions are dual to interface conformal field theories at finite temperature. An approximate  solution is first constructed using perturbation theory around a planar BTZ black hole.  Numerical  and exact solutions valid for all sets of parameters are then found and compared. Using the exact solution the thermodynamics of the system is analyzed. The entropy associated with the Janus black hole is calculated and it is found that
the entropy of the black Janus is the sum of the undeformed black hole entropy and the entanglement
entropy associated with the defect.

\vspace{1.75cm}
\end{titlepage}

\section{Introduction}

The Janus solutions \cite{Bak:2003jk} in various supergravity theories provide interesting  realizations of interface conformal field theories within the AdS/CFT correspondence \cite{Maldacena:1997re,Gubser:1998bc,Witten:1998qj}.
The simplest examples are  constructed using an $AdS_{d}$ slicing of $AdS_{d+1}$ and making a massless field dependent on the slicing coordinate.
Generically the scalar approaches two different constant values at the boundary of the space. Since a massless field is dual to a marginal operator the holographic interpretation is that the coupling constant associated with the marginal operator jumps across the  interface. The nontrivial profile of the massless scalar breaks the full $SO(d,2)$ conformal symmetry but due to the $AdS_{d}$ factor the solution has $SO(d-1,2)$ interface conformal symmetry\footnote{Many generalizations of Janus solutions have been found over the years, see e.g. \cite{D'Hoker:2007xy,D'Hoker:2008wc,Chiodaroli:2009yw,D'Hoker:2007fq,Clark:2005te,Lunin:2007ab,Kumar:2003xi,Chiodaroli:2011nr,Bak:2007jm}.}.

An important feature of the AdS/CFT correspondence is the fact that a black hole in the bulk of the AdS space is dual to a CFT at finite temperature \cite{Witten:1998zw}. It is natural to ask what is the bulk description of an interface CFT at finite temperature, {\sl i.e.} a Janus black hole.  In general this is a complicated problem due to the fact that a (planar) black hole in $AdS_{d+1}$  has translational invariance in the $d-1$ spatial directions. In a Janus solution the nontrivial scalar profile will break this translation invariance. Consequently a generic ansatz for a Janus black hole involves dependence on two spatial coordinates and hence solving nonlinear coupled partial differential equations, where numerical methods  offer  the only approach to solve the problem.

In this paper we will focus on the simplest case of the Janus black hole in three dimensions. The simplification is two fold, both the  BTZ  black hole solution as well as the Janus solution are simpler than in $AdS_{d+1}$ with $d > 2$.
Secondly we find an exact analytical solution\footnote{See also \cite{Bak:2007jm,Bak:2007qw,Bak:2006nh} for the other kinds of closely
related exact solutions.}
for the Janus black hole which can be used to study analytically the physics of the interface CFT as well as test the numerical solution.
Both simplifications are not present in higher dimensions but we believe that, as is often the case in AdS/CFT, there are valuable lessons to be learned from the lower dimensional exactly soluble system.

The structure of the paper is as follows: In section \ref{sec2} we give a brief review of the Janus solution (at zero temperature) and the holographic calculation of the entanglement entropy of the interface CFT. In section \ref{sec3} a small dilaton perturbation with a Janus profile around a planar BTZ black hole is constructed and it is shown that the boundary stress tensor is unchanged to leading order in the perturbation. The numerical solution for the Janus black hole is discussed in section \ref{sec4}.  The main technical problem, that the domain of the coordinates is not fixed but dependent on the solution, is solved by using the dilaton itself as a coordinate.
In section~\ref{sec5} the exact solution  is constructed and
used to test the accuracy of the numerical solution. In section \ref{sec6} we calculate various thermodynamic quantities associated with the black Janus solution. In particular we calculate the entropy, being the horizon length of the black Janus solution, using two methods.
First, we use a boundary-horizon map based on bulk conservation law considerations and second, we use a map which maps
the boundary to the horizon using null geodesics. Interestingly both calculations give the same result and the entropy is the sum of the BTZ black hole entropy and the entanglement entropy of the zero temperature interface CFT. We close with a discussion of our results in section \ref{sec7}. Various  details of calculations are relegated to appendices.

\section{Janus system}\label{sec2}

In this section we will briefly review the construction of the three dimensional
Janus solution \cite{Bak:2003jk,Bak:2007jm} and
its holographic description as an interface conformal field theory (ICFT).

The three dimensional Janus solution can be embedded into ten dimensional type IIB supergravity by
the following ansatz for the metric
\eq
ds^2= e^{\phi/2} (ds_3^2+ ds_{S_3}^2) + e^{-\phi/2} ds_{M_4}^2
\eqx
where $M_4$ is either $T^4$ or $K3$.  For the non supersymmetric Janus solution  the dilaton and metric are independent of the coordinates of the three sphere and $M_4$. All other type IIB  supergravity fields are set to zero except
the three-form field strength $F_3$ \cite{Bak:2007jm}, whose explicit form is not needed in our later discussions.
Upon dimensional reduction of  the ten dimensional action on $M_4 \times S_3$ one arrives at three dimensional action for the metric and $\phi$ given by\footnote{It is pointed out by Eoin \'O Colg\'ain that this three dimensional Janus system
can also be consistently embedded into the eleven dimensional supergravity using the extra dimensional geometry $S_2 \times CY_3$ \cite{Colgain:2010rg}. This 
is dual to an interface version of the
chiral $(4,0)$ SCFT in two dimensions.}
\eq
S={1\over 16\pi G}\int d^3 x \sqrt{g}\left( R- g^{ab}\partial_a \phi \partial_b \phi + 2\right)
\eqx
From this action the Einstein equation  becomes
\eq\label{eqofma}
R_{ab}+2 g_{ab} = \partial_a \phi \partial_b \phi 
\eqx
and the scalar equation of motion  is given by
\eq\label{eqofmb}
\partial_a (\sqrt{g} g^{ab} \partial_b \phi)=0 
\eqx

For the static Janus solution the three dimensional metric  and the dilaton take the following form
\eq\label{threedmet}
ds_3^2= dr^2 + f(r)  {-dt^2 + d{\xi}^2\over \xi^2}, \quad \quad \phi=\phi(r)
\eqx
where  we have set the $AdS_3$ radius to one for convenience. In  \cite{Bak:2007jm}
it was shown that the solution of the equations of motion (\ref{eqofma}) and (\ref{eqofmb}) for this ansatz  is  given by

\eq\label{fforma}
f(r) = {1\over 2}\Big( 1+ \sqrt{1-2 \gamma^2} \cosh(2r)\Big)
\eqx
and
\eq
\phi(r) = \phi_0 + {1\over \sqrt{2}} \log\left( {1+ \sqrt{1-2 \gamma^2} +\sqrt{2}\gamma \tanh(r) \over  1+ \sqrt{1-2 \gamma^2} -\sqrt{2}\gamma \tanh(r)}\right)
\eqx
The Janus solution holographically realizes an ICFT where two CFTs defined on $1+1$ dimensional half spaces are glued together over a $0+1$ dimensional interface.
This can be seen as follows: The conformal boundary of the metric (\ref{threedmet}) has three components as $r\to \pm \infty$ and finite $\xi>0$ we can strip off the $1/\xi^2$ factor and the boundary geometry are two copies of $R\times R_+$ spanned by $t,\xi$.
Note that the dilaton approaches two constant values at the boundaries
\eq
\label{e.phias}
\lim_{r\to \pm \infty} \phi(r) = \phi_0 + {1\over \sqrt{2}} \log\left( {1+ \sqrt{1-2 \gamma^2} \pm\sqrt{2}\gamma \over  1+ \sqrt{1-2 \gamma^2} \mp \sqrt{2}\gamma }\right) \equiv \phi_0 \pm \phi_{as}
\eqx
Without loss of generality we may set $\phi_0=0$. For later use it is also convenient to
express $\gamma$ in terms $\phi_{as}$:
\eq
\label{e.gamma}
\gamma={1\over \sqrt{2}}\tanh \sqrt{2}\phi_{as}
\eqx
The values of the dilaton  on the boundary is dual to a modulus of the two dimensional CFT. The third boundary component is at $\xi=0$ (i.e. the boundary of the $AdS_2$ factor) This defines  the interface where the two half planes are glued together. Hence the dual CFT is an ICFT where two CFTs defined on a half line are at different points in their moduli space.

\subsection{Entanglement entropy}

A useful observable in the ICFT is the entanglement entropy which is defined as follows.  The space  on which the CFT is living is divided into two  domains ${\cal A}$ and ${\cal B}$. The total space of states $H$ is expressed product $H=H_{\cal A}\otimes H_{\cal B}$, where $H_{\cal{A}, \cal{B}}$ is supported on ${\cal A}$ and ${\cal B}$ respectively.
A reduced density matrix can be defined by tracing over all states in ${\cal B}$,
\eq
\rho_{\cal A} = tr_{H_{\cal B}} \rho
\eqx
where $\rho$ is the density matrix of the total system (at zero temperature this is just the projector on the ground state). The entanglement entropy  associated with the domain ${\cal A}$ is then defined as
\eq\label{entangent}
S_{\cal A} =- tr_{H_{\cal A}} \rho_{\cal A} \log \rho_{\cal A}
\eqx

A holographic prescription to calculate the entanglement entropy in  spaces which are asymptotic to $AdS_{d+1}$ was presented  in \cite{Ryu:2006bv,Ryu:2006ef}.
For the domain ${\cal A}$ we denote the boundary $\partial {\cal A}$ which separates it from $\cal B$.
A static minimal surface $\Gamma_{\cal A}$ which extends into the $AdS_{d+1}$ bulk and ends on $\partial {\cal A}$
as one approaches the boundary of $AdS_{d+1}$.
The holographic entanglement entropy can then be calculated as
\eq
S_{\cal A} = {{\rm Area}(\Gamma_{\cal A}) \over 4 G^{(d+1)}}
\eqx
where ${\rm Area}(\Gamma_{\cal A})$ denotes the area of the minimal surface $\Gamma_{\cal A}$ and $ G^{(d+1)}$ is
the Newton constant for $AdS_{d+1}$ gravity.

In the present paper we consider the Janus deformation of $AdS_3$ and the area ${\cal A}$ is an interval and the boundary $\partial {\cal A}$ are the two end points of the interval.
The minimal surface is a space-like geodesic connecting the points.
The geodesic which was used in \cite{Azeyanagi:2007qj} to compute the entanglement    entropy chooses
the $\xi$ coordinate as  constant $ \xi =\xi_{0}$,
while $r$ varies from $- \infty$ to $+ \infty$. This corresponds to a symmetric region  of width $2\xi_{0}$
around the interface.

The geodesic length is divergent and has to be regularized by introducing a cutoff $\epsilon$  near the boundary \cite{Azeyanagi:2007qj}
 \eq
\label{geolength}
{\rm Area}(\Gamma) = R_{AdS_3}\int dr= R_{AdS_3}\big(r_{\infty} (\Gamma) - r_{-\infty} (\Gamma)\big)
\eqx
where the regularized length can be read off from (\ref{fforma})
\eq
r_{\pm \infty} = \mp\Big(\log \epsilon  +{1\over 2} \log  \sqrt{1-2\gamma^2}- \log(2\xi_{0})\Big)
\eqx
Hence
\begin{eqnarray}\label{gfachol}
{\rm Area}(\Gamma) &=&R_{AdS_3}\Big( r_{\infty} (\Gamma) - r_{-\infty} (\Gamma) \Big) \nonumber\\
&=&R_{AdS_3} \Big(  2 \log {2\xi_{0}\over \epsilon} -\log (\sqrt{1-2\gamma^2})\Big)
\end{eqnarray}
The holographic result has the same general form as the entanglement entropy calculated on the CFT side using the replica trick \cite{Calabrese:2004eu},
\eq
S_{\cal A} = {c\over 6} \log {L\over \epsilon} + \log g_{\cal A}
\eqx
where we identify the length of the interval $L=2\xi_{0}$ and  $\epsilon$ is  the UV cutoff.  The last term is boundary entropy (sometimes called  g-factor \cite{Affleck:1991tk})  which is associated with the degrees of freedom localized on the interface. In Ref.~\cite{Azeyanagi:2007qj}  it was shown that in an expansion in the  small deformation parameter $\gamma$ the
holographic result (\ref{gfachol}) agrees to leading order with the weakly-coupled CFT calculation of the boundary
entropy. Note that in the supersymmetric generalization of the Janus solution  \cite{Chiodaroli:2010ur,Chiodaroli:2010mv} one finds exact agreement to all orders.  In Appendix A, we redo the leading order computation  using conformal perturbation theory,
which is  valid for the strongly coupled limit.

\section{Black Janus as a perturbation of the planar BTZ black hole}\label{sec3}

\subsection{Planar BTZ Black Holes}
We shall begin our discussion of three dimensional Janus black holes by studying the leading
order corrections to the  geometry and scalar field starting from the planar BTZ black
hole solution.
The planar BTZ black hole  in three dimensions  \cite{Banados:1992wn} can be written as
\eq
ds^2= \f{1}{z^2} \left[ (1-z^2) d\tau^2+dx^2+\f{dz^2}{1-z^2} \right] 
\label{btz}
\eqx
where we take the range of coordinate $x$ as $(-\infty,\,\infty)$.
Of course the $x$ direction may be compactified on a circle  but we shall be concerned
here only with the non compact case.  The horizon is located at $z=1$ and a convenient
change of variables for checking its regularity is
\eq
\tilde{z}^2= 1-z^2 
\eqx
In this coordinate system the black hole looks like
\eq
ds^2=\f{1}{1-{\tilde{z}}^2} \left[ {\tilde{z}}^2 d\tau^2+dx^2+\f{d{\tilde{z}}^2}{1-{\tilde{z}}^2} \right]\,,
\eqx
and now the AdS boundary is located at ${\tilde{z}}=1$ while the horizon at ${\tilde{z}}=0$.
Since
\eq
\f{d{\tilde{z}}^2}{1-{\tilde{z}}^2}+ {\tilde{z}}^2 d\tau^2 \sim d{\tilde{z}}^2+{\tilde{z}}^2 d\tau^2\,
\eqx
in the near horizon regime, there is no conical singularity if the euclidean time coordinate $\tau$ is periodic with period $2\pi$. Therefore the corresponding temperature can be identified as
\eq
T={1\over 2\pi} 
\eqx
The BTZ black hole with general temperature is described
by the metric
\eq
ds^2= \f{1}{{z'}^2} \left[ (1-a^2\,{z'}^2)
d{\tau'}^2+d{x'}^2+\f{d{z'}^2}{1-a^2\,{z'}^2} \right] 
\eqx
which can be obtained by the scale coordinate transformation
\eq
z'=a z 
\ \ \ \ \ \tau'=a \tau 
\ \ \ \ \  x'=a x
\eqx
from (\ref{btz}).
The temperature for this scaled version of the black hole now becomes
\eq
T'= {a\over 2\pi } 
\eqx
Below we  work mostly with the temperature $T=(2\pi)^{-1}$ and,
using the above freedom of scale transformation, we shall recover the general temperature dependence
whenever it is necessary.

\subsection{Linearized Black Janus}

Introducing a new coordinate $y$ given
by $z=\sin y$, the planar black hole metric (\ref{btz}) can be rewritten as
\eq
ds^2=\f{1}{\sin^2 y} \left[\, \cos^2 y \, d\tau^2+dx^2+dy^2 \right] 
\eqx
Motivated by the form of this metric, we shall make the following ansatz for the black Janus solution
\eq
ds^2=  {dx^2 + dy^2\over A(x,y)}+{d\tau^2\over B(x,y)}
\quad \quad \quad \phi=\phi(x,y)
\eqx
It is then straightforward to show that the equations of motion (\ref{eqofma}) and (\ref{eqofmb})  reduce to
\eqn
&&({\vec\partial} A)^2 -A\, {\vec\partial}^2 A=2A - A^2 \,({\vec\partial} \phi)^2\\
&& 3({\vec\partial} B)^2 -2 B\, {\vec\partial}^2 B=8 B^2/A \\
&& {\vec\partial} B \cdot {\vec\partial} \phi -2 B \, {\vec\partial}^2 \phi=0
\eqnx
where we introduced the notation
${\vec\partial}=(\partial_{x},\partial_{y})$.
To the leading order, the scalar equation can be integrated as
\eq
\label{e.scalarlin}
\phi=\gamma\, \f{\sinh x}{\sqrt{\sinh^2 x +\sin^2 y} }+ O(\gamma^3)
\eqx
where we have used the Janus boundary condition
$\phi(x,0)= \gamma\,  {\epsilon}(x)+O(\gamma^3)$ with the sign function $\epsilon(x)$. The leading perturbation
of the metric part is of order $\gamma^2$. For which we set
\eq
A=A_0 \Big(1+{\gamma^2\over 4}a(x,y)+O(\gamma^4)\Big)
\ \  \ \ \
B=B_0 \Big(1+{\gamma^2\over 4}b(x,y)+O(\gamma^4)\Big)
\eqx
with
\eq
A_0 =\sin^2 y
\ \ \ \ \  B_0 = \tan^2 y 
\eqx
The leading order equations for the metric part becomes
\eqn
&& 2 a -\sin^2y\, {\vec\partial}^2 a
=-4\f{\sin^4y}{(\sinh^2 x+\sin^2 y)^2}
\label{eq1}
\\
&&  2 \tan y \,{\partial_y} b -\sin^2 y\, {\vec\partial}^2 b +4 a=0 
\label{eq2}
\eqnx
With the Janus boundary condition whose detailed structure will be discussed later on, the
solution can be found as $a(x,y)=b(x,y)=q(x,y)$ where
 \eq\label{qsolu}
q(x,y)= 3 \left(\f{\sinh x}{\sin y}\right) \,\tan^{-1}\left(\f{\sinh x}{\sin y}\right)
+\f{\sinh^2 x}{\sinh^2 x+\sin^2 y}+2 +(c_1-c_2) {\sinh x \over \sin y} 
\eqx
with $c_1$ and $c_2$ being $O(1)$ integration constants\footnote{Only the combination $c_1-c_2$ is a true
integration constant.
We break it up to two since their roles are different in the discussions below.}.
Indeed checking that (\ref{qsolu}) solves
eqs. (\ref{eq1}-\ref{eq2}) is straightforward.
Then the metric for the black Janus can be written as
\eq
ds^2=  {1-\f{\gamma^2}{4} q(x,y)\over \sin^2 y}\left[\, \cos^2 y \, d\tau^2+dx^2+dy^2 \right]+O(\gamma^4)
\label{bmetric0}
\eqx
Next we  introduce  a new  angular coordinate $\mu$ that is defined by
\eq
\tan \Big(\mu+ {\gamma^2\over 4} c_1 \Big) = \f{\sinh x}{\sin y} 
\eqx
The above metric for the linearized black Janus can be written using the scale function $f(\mu)$
of the original Janus solution: Namely,
 the metric can be expressed in the following form
\eqn
 ds^2=\f{f\Big(\mu+ {\gamma^2\over 4} c_2 \Big)}{\sinh^2 x+\sin^2 y} \left[\, \cos^2 y \, d\tau^2+dx^2+dy^2 \right]+O(\gamma^4) 
\label{bmetric}
\eqnx
where
\eq
\label{e.fmu}
f(\mu)=\f{\kappa_+^2}{{\rm sn}^2 (\kappa_+ (\mu+\mu_0),k^2)}
\eqx
with
\eqn
&&\kappa^2_\pm \equiv \f{1}{2} (1\pm\sqrt{1-2\gamma^2})\\
&& k^2 \equiv \kappa^2_-/\kappa^2_+={\gamma^2\over 2}+O(\gamma^4)\\
&& \mu_0 \equiv K(k^2)/\kappa_+ =\f{\pi}{2} \left(1+ \f{3}{8}\gamma^2+ O(\gamma^4)\right) 
\eqnx
To show this, we have used the expansion of the scale function $f(\mu)$ given in the form
\eq
f\Big(\mu+ {\gamma^2\over 4} c_2 \Big)={1-\f{\gamma^2}{4} q(x,y)\over \cos^2 \Big(\mu+ {\gamma^2\over 4} c_1 \Big)}
+O(\gamma^4) 
\eqx
As will be explicitly verified later on, the remaining part of the metric except
the scale factor $f$ possesses  a translational  isometric direction along $\mu$. Hence
one may set $c_2=0$ without loss of generality.   The zeroes of the function $A$ and $B$ occur at $\mu=\pm \mu_0$,
which correspond to the boundary of the asymptotically AdS space. As a consequence the coordinate $\mu$ is ranging
over $[-\mu_0,\, \mu_0]$ as the case of the original Janus solution.

The choice of the integration constant $c_1$ is simply related to the
the choice of the coordinate patch of $(t,x,y)$, which covers only part of the entire black hole geometry.
Let us first consider the choice $c_1=0$. Then the boundary occurs at
\eq
{\sin y\over |\sinh x|}= \tan \Big({\pi\over 2}-\mu_0\Big) +O(\gamma^4) 
\eqx
This is then solved by
\eq
\sin y(x) = -{3\pi \over 16}\gamma^2 |\sinh x|+ O(\gamma^4) 
\label{yb}
\eqx
Therefore  the validity of the coordinate is limited by
\eq
3\pi \gamma^2 |\sinh x|/16 \le 1  
\eqx
and, hence, the coordinate along the boundary become singular if
\eq
|x| \ > \ x_{\rm cut} \sim \ln (2/\gamma^2) 
\eqx
Only the region $-{\pi\over 2}\le \mu \le {\pi\over 2} $ can be free of any such coordinate problem.
But this is simply a coordinate singularity, which may be removed  by choosing a different
coordinate chart. For instance let us consider the choice ${\gamma^2\over 4} c_1= {\pi\over 2} -\mu_0 +O(\gamma^4)$.
For this case, one boundary can be solved by ${\sin y\over \sinh x} = 0$ for $x\  > \ 0$, whose solution is
$y=0$. For this side we do not have any coordinate problem but the other side of the boundary has again the
coordinate singularity. The region specified by $\mu_0 -\pi\le \mu \le \mu_0$ of this coordinate chart does not
involve any coordinate problem. By the choice of ${\gamma^2\over 4} c_1= -{\pi\over 2} +\mu_0 +O(\gamma^4)$,
the $-\mu_0\le \ \mu \ \le -\mu_0+\pi$ region can be safely covered, which includes the
other side of the boundary.

For our further analysis of geometry below, we shall simply choose $c_1=0$  (together
with $c_2=0$), since the presence of the coordinate singularity can be
ignored in the small $\gamma$ limit.

\subsection{Boundary stress tensor}
In this subsection, we shall construct the Fefferman-Graham  metric
to determine
the boundary stress energy tensor. In order to use the prescription developed in
Ref.~\cite{de Haro:2000xn},
we introduce the metric in the
following Fefferman-Graham form,
\eq
ds^2={d\chi^2\over \chi^2}  +{1\over \chi^2} g_{\mu\nu}(X,\chi) dX^\mu dX^\nu
\label{fgform}
\eqx
where $X^\mu$ ($\mu=0,\,1$) denote the boundary coordinates\footnote{We use this notation of boundary coordinates
for this subsection only. In the subsequent (sub)sections we shall simply use $t$ and $x$ for the boundary coordinates for the
notational simplicity.
} and $\chi=0$ corresponds to
the location of the boundary.
In general one may expand $g_{\mu\nu}$ by
\eq
g_{\mu\nu}(X,\chi)=g^{(0)}_{\mu\nu}(X) + \chi^2 g^{(2)}_{\mu\nu}(X)+\cdots
\eqx
where $g^{(0)}_{\mu\nu}$ is the metric for the boundary system.
In three dimensions, the boundary stress energy tensor is then given by \cite{de Haro:2000xn}
\eq
T_{\mu\nu}(X) = {1\over 8\pi G}
\left[g^{(2)}_{\mu\nu}(X)-g^{(0)}_{\mu\nu}(X)\,g^{(2)}_{\alpha\beta}(X) g^{(0)\alpha\beta}(X)\right]+
\tau_{\mu\nu}(X) 
\eqx
where $\tau_{\mu\nu}(X)$ is the scalar contribution for the stress energy tensor
given by
\eq
\tau_{\mu\nu}(X)={1\over 8\pi G}\Big[\partial_\mu \phi_B\partial_\nu \phi_B
- {g^{(0)}_{\mu\nu}\over 2}\, g^{(0)\alpha\beta}\partial_\alpha \phi_B\partial_\beta \phi_B \Big] 
\eqx
with $\phi_B$ denoting the boundary value of the scalar field.
For our case, the boundary metric is given by
\eq
g^{(0)}_{\mu\nu}={\rm diag}(-1,1)=\eta_{\mu\nu} 
\eqx
since the boundary system is defined in the flat Minkowski space
in two dimensions and the scalar contribution to the stress energy tensor vanishes
since the scalar field is constant except $X_1=0$.
Let us first bring the metric in (\ref{bmetric})
to the form
\eq
ds^2= {dY^2\over \sin^2 Y} +{dX_1^2\over \sin^2 Y}\Big(1-{\gamma^2\over 4} C\,\Big)
-{dX_0^2} \cot^2 Y\, \Big(1-{\gamma^2\over 4} D\,\Big)
\eqx
where $X_0=i\tau$.
Introducing ${\cal X}(x,y)$ and ${\cal Y}(x,y)$ by
\eqn
X_1(x,y)=x-{{\gamma^2\over 8}} {\cal X}(x,y)+O(\gamma^4)
\ \  \ \ \ Y(x,y)= y -{{\gamma^2\over 8}} {\cal Y}(x,y)+O(\gamma^4) 
\eqnx
and comparing the two forms of the  metric to the leading order of $\gamma^2$, one finds the
differential equations,
\eqn
 \partial_y  {\cal Y} -  {\cal Y}\,\cot y = q(x,y) 
 \ \ \ \ \
\partial_x  {\cal Y} +\partial_y  {\cal X}=0 
\eqnx
with
\eqn
&& C(x,y)=q(x,y)- \partial_x  {\cal X} +  {\cal Y}\, \cot y 
\nonumber\\
&& D(x,y)=q(x,y)+{ {\cal Y}\over \sin y\cos y} 
\eqnx
The boundary conditions
$C(x,0)=D(x,0)=0$
are required to have the standard form of the boundary metric
$\eta_{\mu\nu}$.
The solution satisfying the boundary conditions is uniquely found by
\eqn
 {\cal X}(x,y)&=& 3\sin y \cosh x \tan^{-1}{\sinh x\over \sin y} +\big(1-\cos y\big)\,{3\sinh^2 x+2\over \sinh x \cosh x}\nonumber\\
&-&\cos y \sinh x {3\cosh^2 x+1\over \cosh^2 x}\Big[\tanh^{-1}{\cos y\over \cosh x}
-\tanh^{-1}{1\over \cosh x}
\Big]\nonumber\\
 {\cal Y}(x,y)&=& -3\cos y \sinh x \tan^{-1}{\sinh x\over \sin y}-3\sin y
\nonumber\\
&-&\sin y {3\sinh^2 x+2\over \cosh x}\Big[\tanh^{-1}{\cos y\over \cosh x}
-\tanh^{-1}{1\over \cosh x}\Big] 
\eqnx
with
\eqn
C(x,y)&=&  {2\cos y\over \cosh^3 x}\Big[\tanh^{-1}{\cos y\over \cosh x}
-\tanh^{-1}{1\over \cosh x}\Big]
\nonumber\\
&+&{2(1-\cos y)\over \sinh^2 x\cosh^2 x}\nonumber\\
D(x,y)&=&-{\sec} y {3\sinh^2x+2\over \cosh x}\Big[\tanh^{-1}{\cos y\over \cosh x}
-\tanh^{-1}{1\over \cosh x}\Big]
\nonumber\\
&-&{\sin^2 y\over \sinh^2 x +\sin^2 y}+3(1-{\sec} y) 
\eqnx
One may check that the location of boundary $y(x)$ in (\ref{yb}) corresponds to
$Y=0$ as expected.

For the stress energy tensor, we now note that
$C(x,y)=D(x,y)=O(y^4)$,
which implies that the $O(\gamma^2)$ terms have no contribution to the stress energy tensor.
The rest is then straightforward:
Noting
\eq
\chi=2\tan {Y\over 2} 
\eqx
one finds
\eq
g_{\mu\nu}^{(2)}={1\over 2}{\rm diag}(+1,+1)+O(\gamma^4) 
\eqx
Therefore, one has
\eq
T_{\mu\nu} ={1\over 16\pi G} {\rm diag}(+1,+1)+O(\gamma^4) 
\eqx
Hence to the leading order in $\gamma$, the stress energy tensor is independent of the
deformation.
Later on we shall show that
$T_{\mu\nu}$ is in fact $\gamma$ independent and the zeroth order result
is all order exact.
Finally recovering the temperature dependence by the scaling transformation, we have
\eq
T_{\mu\nu}= {\pi \, T^2\over 4 G} {\rm diag}(+1,+1) 
\label{stress}
\eqx
which agrees with that for the usual BTZ black hole.

\section{Black Janus at arbitrary $\gm$ --- Numerical Approach}\label{sec4}

In order to obtain the form of the Janus black hole in the fully nonlinear
regime of arbitrary $\gm$ we develop a numerical approach to solving the relevant
Einstein+dilaton system of equations. It turns out that a successful implementation
is surprisingly subtle, due to the special features of the Janus system.

The most naive guess for the numerical ansatz would be
\eq
ds^2= \f{1}{\sin^2 y} \left[ e^{V(x,y)+W(x,y)} \cos^2 y\, d\tau^2+dx^2+
e^{V(x,y)-W(x,y)} dy^2 \right]
\eqx
with the dilaton given by $\phi(x,y)$. This form of the metric ansatz makes it very
easy to implement both constant temperature (which corresponds to Dirichlet boundary conditions
for $W(x,y)$ at $y=\pi/2$ and Neumann for $V(x,y)$) and the Janus boundary condition.
For numerics, we should map the infinite spatial coordinate into a finite interval
e.g. by the mapping $s=\tanh x$.
However, this metric leads to several problems, some
purely numerical and some, what is more dangerous, conceptual. Firstly, the discontinuous
boundary condition for the dilaton with the jump at $x=0$ is very difficult to handle
numerically. Secondly, it is far from clear what would be the domain of definition
of the exact solution. The range of the bulk $y$ coordinate is $y\in [0\,,\pi/2]$, however
the range of $s$ is unknown. At the boundary the range is $s\in [-1\,, 1]$, but in the bulk
it may well be $s\in [-s_{max}(y)\,, s_{max}(y)]$ with an \emph{a-priori} unknown
profile $s_{max}(y)$.
This indeterminacy \emph{a-priori} precludes any numerical treatment.

In order to overcome the above difficulty, it is convenient to link the spatial
coordinate to the value of the dilaton, since then the asymptotic range of the
spatial coordinate is fixed by the definition of the Janus system, since at spatial
infinities the dilaton is constant in the bulk and attains its asymptotic value $\pm \phi_{as}$.
This leads to the following ansatz (recall (\ref{e.phias}))
\eqn
ds^2&=&\f{d\tau^2}{\tan^2 y} +A(y,s) dy^2+2B(y,s) dy ds+C(y,s) ds^2\\
\phi&=&\phi_{as} \, s 
\label{e.defphis}
\eqnx
We have thus to deal with a non diagonal metric. Moreover it is not obvious what
should be the boundary conditions characteristic of asymptotically AdS spacetime.
To this end let us consider as a first approximation
the BTZ black hole and use the linearized dilaton perturbation to fix the spatial
coordinate according to (\ref{e.defphis}).

The standard BTZ black hole metric is
\eq
ds^2= \f{d\tau^2}{\tan^2 y}+ \f{1}{\sin^2 y} \left( dy^2+ \f{d\st^2}{
(1-\st^2)^2} \right)
\eqx
where we introduced $\st=\tanh x$. Now the linearized Janus perturbation
(\ref{e.scalarlin}) takes the form
\eq
\phi=\f{\phi_{as} \st}{\sqrt{\st^2+(1-\st^2)\sin^2 y}}
\eqx
which leads to the following change of coordinates
\eq
\st=\f{s \sin y}{\sqrt{1-s^2 \cos^2 y}}
\eqx
In these `dilaton-adjusted' $s-y$ coordinates the BTZ black hole takes
the following form
\eq
ds^2=\f{d\tau^2}{\tan^2 y}+ \f{1}{1-s^2 \cos^2 y} \left[
\f{dy^2}{\sin^2 y} + \f{2 s\cos y\, ds\, dy}{\sin y (1-s^2)} +\f{ds^2}{(1-s^2)^2}
\right]
\eqx
which suggests the following ansatz for numerical computations
\eq
ds^2=\f{d\tau^2}{\tan^2 y}+ \f{1}{1-s^2 \cos^2 y} \left[
\f{\tilde{K}(y,s)dy^2}{\sin^2 y} + \f{2 \tilde{L}(y,s)\, ds\, dy}{\sin y (1-s^2)} +
\f{\tilde{M}(y,s)ds^2}{(1-s^2)^2} \right]
\eqx
with very smooth solutions for $\phi_{as}=0$: $\tilde{K}=\tilde{M}=1$ and $L=s\cos y$.
Performing numerics with such an ansatz shows that even for very small $\gm$
(equivalently $\phi_{as}$), the coefficient functions have different limits as
$y\to 0$ with $s=1$ fixed and as $s\to 1$ keeping $y=0$ fixed. This leads to severe
numerical problems and indicates that the pre-factor
\eq
\f{1}{1-s^2 \cos^2 y}
\eqx
should be replaced by a suitable $\gm$-dependent function.

Fortunately, the Einstein-dilaton equations for the coefficient functions at $y=0$
reduce to \emph{ordinary differential equations} with no $y$-derivatives, which can be
solved exactly. Particularly relevant is the solution for $\tilde{K}(0,s)$:
\eq
\label{e.ktilde}
\tilde{K}(0,s)= \al^2 \f{\phi_{as} (1-s^2) \sinh \sqrt{2}\phi_{as}}{
\sqrt{2}(\cosh\sqrt{2}\phi_{as}-\cosh \sqrt{2}s)}
\eqx
with
\eq
\al^2=\f{\tanh\sqrt{2}\phi_{as}}{\sqrt{2}\phi_{as}}
\eqx
Taking into account the asymptotic BTZ metric at spatial infinity
\eq
\tilde{K}(y,1)=\tilde{M}(y,1)=1 \qqqq \tilde{L}(y,1)=\cos y
\eqx
we are led to modify the pre-factor
\eq
\f{1}{1-s^2 \cos^2 y} \to f(y,s)
\eqx
to take into account these properties, namely we require that
\eqn
f(0,s) &=& \f{\al^2}{1-s^2} \\
f(y,1) &=& \f{1}{\sin^2 y}
\eqnx
In addition it is convenient to have $f(\pi/2,s)=1$ and
$\partial_y f(\pi/2,s)=0$ so as not to modify the form of boundary conditions
at the horizon. A function which satisfies all the above properties is
\eq
f(y,s)=\f{\al^2+(1-\al^2)(1-s^2)\sin^2 y}{1-s^2(1-\al^2 \sin^2 y)}
\eqx

This leads us to the final ansatz for the numerical solution:
\eq
\label{e.finalansatz}
ds^2=\f{d\tau^2}{\tan^2 y}+ f(y,s) \left[
\f{e^{K(y,s)}dy^2}{\sin^2 y} + \f{2 L(y,s)\, ds dy}{\sin y (1-s^2)} +
\f{e^{M(y,s)}ds^2}{(1-s^2)^2} \right]
\eqx
At the horizon $y=\pi/2$ we impose the following boundary conditions:
\eq
K(\f{\pi}{2},s) = 0 \qqqq
L(\f{\pi}{2},s) = 0 \qqqq
\partial_y M(\f{\pi}{2},s) = 0
\eqx
The first condition ensures that the temperature is constant.
At $s=0$ we impose boundary conditions following from symmetry
\eq
\partial_s K(y,0) = 0 \qqqq
L(y,0) = 0 \qqqq
\partial_s M(y,0) = 0
\eqx

\subsection{Numerical details}

A feature of the Einstein's equations for the coordinate ansatz
(\ref{e.finalansatz}) is that
we can pick three independent \emph{first order} equations. These are the
equation for $\phi$, the $\tau\tau$ component of Einstein's equations $E_{\tau\tau}$,
and an appropriate linear combination of $E_{yy}$ and $E_{ss}$. Subsequently we
solve the above three equations numerically.

We use a Chebyshev grid of $N=20$ or $N=30$ points in each dimension and use
the PETSc library
for solving nonlinear equations. We use a Python interface to the library ({\tt petsc4py}).
Unfortunately the convergence is poor so we had to use an automatic differentiation
package (ADOL-C with Python bindings {\tt pyadolc}) to compute the Jacobian and use
LU linear solver from PETSc instead of the standard iterative ones.
In addition the numerical solutions were found by gradually increasing
the asymptotic value of the dilaton $\phi_{as}$ by $0.05$ from the BTZ value
of $\phi_{as}=0$ and using the output
from the previous value of $\phi_{as}$ as initial conditions for $\phi_{as}+0.05$. In this way we generated
the metric profiles up to $\phi_{as}=10.0$.

The above numerical setup should be readily generalizable to the Janus black holes
in higher number of dimensions for which, almost certainly an analytical solution does not
exist.

\section{Black Janus at arbitrary $\gm$ --- Exact solution in 3D}\label{sec5}

As an alternative to the direct numerical solutions, one may attempt to use
the above properties of the metric, such as the form of (\ref{e.ktilde}) to try
to obtain an exact analytical solution. Remarkably this can be done for the
three dimensional  black Janus system considered in this paper.

\begin{figure}[t]
\hfill
\includegraphics[height=6cm]{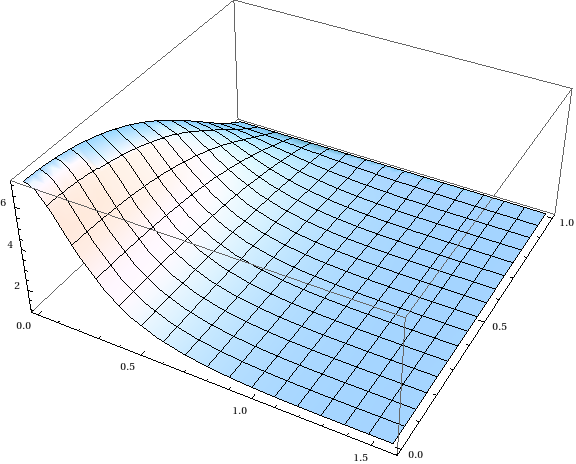}
\hfill
\includegraphics[height=6cm]{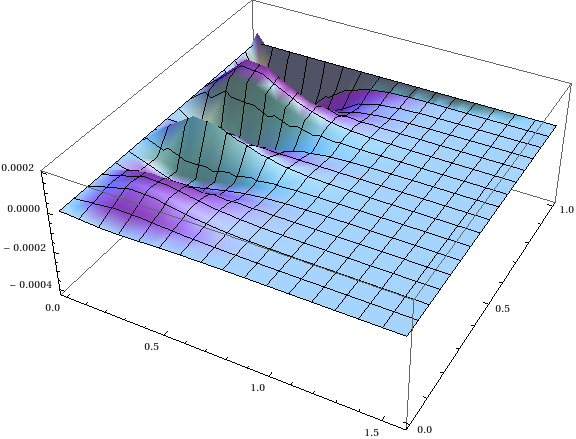}
\hfill
\caption{On the left hand side we show $e^{K(y,s)}$ for the numerical solution
with $\phi_{as}=10.0$, and on the right its relative deviation from the exact
analytical solution.}
\label{f.numerics}
\end{figure}

The \emph{exact} Black Janus solution for arbitrary $\gm$ turns out to be given by the
following analytical expression:
\eqn
\label{e.exact}
&&ds^2=\cot^2 u d\tau^2
+F(u,\varphi)\Big[\,{du^2\over \sin^2u}
+\cot u\, (\log f(\varphi))'du d\varphi
\nonumber\\
&&+{\phi^2_{as}\,f^2(\varphi)\over \gamma^4}
\Big(\gamma^2\sin^2 u +
\cos^2 u \Big[
1\!-\!{\cosh \sqrt{2}\phi_{as}\varphi\over \cosh \sqrt{2}\phi_{as}}
\!+\!{\sinh^2 \sqrt{2}\phi_{as}\varphi\over 2\cosh^2 \sqrt{2}\phi_{as}}
\Big]
\Big)
d\varphi^2
\Big]
\eqnx
where
\eq
F(u,\varphi)=\Big[
\sin^2u+{\cos^2u \over f(\varphi)}
\Big]^{-1} 
\eqx
and
\eq
f(\varphi)
={\gamma^2\over 1-{\cosh \sqrt{2}\phi_{as}\varphi\over \cosh \sqrt{2}\phi_{as}} }
\eqx
In this form we see the way that the exact metric incorporates the behavior
(\ref{e.ktilde}). The similarity with the numerical ansatz made it easy to
compare the above exact expression with the numerical solution for quite large
value of $\phi_{as}=10.0$ with excellent agreement. In figure~\ref{f.numerics}
we show the numerically obtained $e^{K(y,s)}$ coefficient function from (\ref{e.finalansatz})
together with the relative deviation from the exact solution obtained from (\ref{e.exact}).

It turns out, however, that passing to another coordinate system allows us to
drastically simplify the metric. Indeed, let us change the $u$ coordinate
into $w$ through the expression
\eq
\cot^2 u= f(\varphi) (w^2-1) 
\eqx
Then the exact metric transforms into
\eq
ds^2=f(\varphi) \left[ ( w^2-1) d\tau^2+ \f{dw^2}{w^2-1} +{\phi^2_{as}\over \gamma^2} f(\varphi)
d\varphi^2 \right]
\eqx
This can be further simplified introducing the coordinate $p$ instead of $w$ through
$w=\cosh p$. With this substitution, we obtain a remarkably simple form of
the exact Janus black hole metric:
\eqn
ds^2&=&f(\varphi)\Big[
\sinh^2 p\, d\tau^2 +{dp^2}+{\phi^2_{as}\over \gamma^2} f(\varphi)d\varphi^2
\Big]\nonumber\\
&=&f(\mu)\Big[
\sinh^2 p\, d\tau^2 +{dp^2}+d\mu^2
\Big]
\label{emetric}
\eqnx
where we have used the fact $
\left(d\varphi\over d\mu\right) \sqrt{f(\mu)}=\gamma/\phi_{as}$
with $f(\mu)$ given by (\ref{e.fmu}).

\section{Thermodynamics and Entropy}\label{sec6}

\subsection{Thermodynamic quantities}

In this section, we shall be describing thermodynamic properties
of our Janus system based on the solution of the previous section.
First we turn to  the stress energy tensor. We have already discussed
the stress energy tensor for the case of the linearized black Janus and
claimed that the expression in (\ref{stress}) is in fact exact to all
orders in $\gamma$. Let us argue this point first. From the exact metric
we note that the deformation parameter $\gamma$ dependence occurs
only through the function $f(\mu)$. In the near boundary region
of $\mu \sim \pm \mu_0$, the scale function $f(\mu)$ can be expanded as
\eq
f(\mu)= {1\over (\mu-\mu_0)^2} \left[
1+ {1\over 3} (\mu-\mu_0)^2 +
b_4(\gamma) (\mu-\mu_0)^4 +b_6(\gamma) (\mu-\mu_0)^6+\cdots
\right]
\eqx
and $\gamma$ dependence can only appear in the higher order coefficients
$b_{2n}$ with $n\ge 2$. From dimensional analysis it follows that $(\mu-\mu_0)^{2n}
\sim  {\chi^{2n}/ x^{2n}}$ in the near boundary region where $\chi$ is the
Fefferman-Graham coordinate
introduced in (\ref{fgform}). Hence
$g^{(0)}_{\mu\nu}$ and $g^{(2)}_{\mu\nu}$ are
 $\gamma$ independent since its dependence enters only at higher orders of $\chi$.
 Therefore we conclude that the stress energy tensor in (\ref{stress}) is all
 order exact.

One can reach the same conclusions purely from the field theory perspective.
Consider the energy-momentum conservation equations:
\eq
\partial_t T^{tt}+\partial_x T^{xt} =0 \qqqq
\partial_t T^{xt}+\partial_x T^{xx} =0
\eqx
Due to time reversal invariance we have $T^{xt}=0$. Then $T^{xx}$ is $x$
independent and, due to tracelessness, so is $T^{tt}=0$. So $T^{\mu\nu}$
is diagonal and constant in space and time. For an infinite system that we consider
explicitly here, the value of the energy density equals the value of the energy
density in the asymptotic region and so equals the BTZ value.

Of course our expression of stress tensor is strictly
valid only when the size of the boundary system $L$ goes to infinity.
Dealing with the finite size system, which in general involves the finite size effect,
is not a simple matter. According to the recent proposal \cite{Takayanagi:2011zk},
the boundary $\partial B$ of a boundary system,
can be dealt with in a rather simple manner by introducing corresponding
bulk boundaries $\partial M$.
This hypersurface is extended into the bulk  in an appropriate manner
from $\partial B$
and introduces an extra contribution to
the thermodynamic quantities.
This formalism, however, is developed only for  boundary conformal field
theories and it is not clear  whether it is applicable
to our case or not. Thus in this note we shall be only concerning about
the limit where the system size $L$ goes to
infinity.

In the next two subsections, we shall be calculating the entropy of the black Janus
system utilizing two different methods. Since the translational symmetry in the $x$ direction is broken by the
Janus deformation, the entropy density should be position dependent.
The entropy is a quantity defined at the horizon,  whereas  the interface
CFT and its  stress tensor is naturally defined at the boundary side
of the geometry.
Hence in order to talk about the entropy, one has to relate the horizon
side to the boundary side. Especially one needs a map which relates the boundary
coordinate $x$ to the horizon coordinate in order to use the data found at
the horizon of the black hole. Below we shall discuss two methods for
the boundary-horizon map, which will be used to determine the
entropy of the boundary system.
We shall find that the both methods lead to the desired entropy
in the large $L$ limit to all orders in $\gamma$. Using the map, one may wonder
whether one can define an entropy density by evaluating the corresponding
horizon length scale divided by $4G$. 
The resulting
expressions for the entropy density by the two methods
turn out not to agree with each other even in the
large $L$ limit.  This, however, is not a problem.
As discussed in \cite{Bhattacharyya:2008xc}, the entropy density alone
is not fully well defined in a gauge invariant manner even in the semiclassical limit.
We shall get back to this issue later in the discussion section.

\subsection{Method 1}
\begin{figure}[ht!]
\centering
\includegraphics[width=8cm]{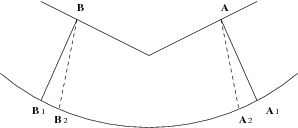}
\caption{\small The finite temperature Janus system, whose spatial extent
is given by the segment $AB$, is corresponding to the bulk
region $AA_1B_1A$. The dotted lines describe the identification for the undeformed
geometry with $\gamma=0$. }
\label{fig1}
\end{figure}

As stated previously,
the entropy is a  quantity  which is defined at the horizon
whereas the interface conformal field theory is defined at
the boundary  of the geometry.
The stress energy tensor is also determined by the behavior of the geometry near the boundary.
The question is how to connect a boundary region given by $AB$ to the corresponding  horizon
region $A_1B_1$ as depicted in Figure~\ref{fig1}.
In other words the question is how to find the two bulk lines $AA_1$ and $BB_1$
which join the image at the horizon to the boundary region (of a coordinate size $L$).
This may be done as follows:
Consider the  conserved current \cite{waldiyer}
\eq
Q_a=\epsilon_{abc} \nabla^b \xi^c
\eqx
where $\xi^c$ is the time translation Killing vector satisfying
the Killing equation $\nabla_a \xi_b +\nabla_b\xi_a=0$.
In our case, $\xi^a =\delta^{a0}$ and one may find the
boundary lines from the condition
\eq
dx^a Q_a =0
\eqx
with $t=constant$.
This basically says that there is no contribution from the bulk lines
for the integral of $Q_a$ over the region surrounded by $AA_1B_1B$.
This condition then ensures the first law of thermodynamics that connects
the change of entropy to the energy and the length of the system.

For the analysis of the exact geometry, we shall use the metric
in (\ref{emetric}).
As in the previous section,
we shall consider the boundary system of spatial size $L$ with the interface
at the center. (Namely $x\in [-L/2, L/2]$ with $x$ as the standard boundary coordinate.)
Following above prescription, the corresponding bulk lines  satisfy
\eq
{d\varphi\over dp}={g_{pp}\, \partial_\varphi g_{tt}
 \over g_{\varphi\varphi}\, \partial_p g_{tt}}\,.
\eqx
This is solved by
\eq
{\tanh \Big({\phi_{as}\varphi\over \sqrt{2}}\Big)
\over \tanh \Big({\phi_{as}\over \sqrt{2}} \Big)} =\Big({\cosh p\over \cosh p_A}\Big)^{1\over \cosh\sqrt{2}\phi_{as}} 
\eqx
For the line starting from $x_A=L/2$,
\eq
\label{e.bndryid}
{1\over \cosh p_A}=\tanh {L\over 2} 
\eqx
This identification of the boundary coordinate for exact black Janus solution is carried out
in Appendix C.
The corresponding horizon point $A_1$ is described by
\eq
\label{mapmethod1}
 \tanh \Big({\phi_{as}\varphi_{A_1}\over \sqrt{2}}\Big)=\tanh \Big({\phi_{as}\over \sqrt{2}} \Big) \,
 \Big(\tanh {L\over 2}\Big)^{1\over \cosh\sqrt{2}\phi_{as}} 
\eqx
together with $p_{A_1}=0$. This is exactly the boundary-horizon map obtained
by the present method.
Then the  horizon length $L_H$ can be obtained by
\eq\label{lhresult}
{L_H\over 2}= {\phi_{as}\over \gamma}\,\int^{\varphi_{A_1}}_0 f(\varphi)d\varphi
={1\over 2}\log \left[
{1+ \Big(\tanh {L\over 2}\Big)^{\sqrt{1-2\gamma^2}}\over
1-\Big(\tanh {L\over 2}\Big)^{\sqrt{1-2\gamma^2}}
}
\right] 
\eqx
where we used
\eq
{1\over \cosh\sqrt{2}\phi_{as}}=\sqrt{1-2\gamma^2} 
\eqx
The entropy then becomes
\eq\label{slresult}
S_L={1\over 4G}\log \left[
{1+ \Big(\tanh {L\over 2}\Big)^{\sqrt{1-2\gamma^2}}\over
1-\Big(\tanh {L\over 2}\Big)^{\sqrt{1-2\gamma^2}}
}
\right] 
\eqx
By expanding in $\gamma^2$, one may find the leading order behavior
as
\eq
\Delta S ={1\over 4 G}(L_H-L)=-{\gamma^2\over 8G} \sinh L \log \tanh^2 {L\over 2} +O(\gamma^4)\rightarrow
{\gamma^2\over 4G}+O(\gamma^4) 
\label{linear}
\eqx
For the large size limit the finite part of  (\ref{slresult}) becomes
\eq
\Delta S_{L=\infty}={1\over 4G}\log (1/\sqrt{1-2\gamma^2}) 
\eqx
This expression precisely coincides with the result from the method of the
 holographic entanglement entropy.

\subsection{Method 2: a boundary horizon map based on null geodesics}
In this subsection, we shall carry out the boundary horizon map based on the
null geodesics emanating from the boundary in a hypersurface orthogonal manner.
This construction is based on the concept of the light-sheet of the holography \cite{Bousso:1999xy}
and widely used in application of fluid-geometry correspondence \cite{Bhattacharyya:2008xc,hol2,hol3,hol4}.

Since null geodesics are the same for conformally equivalent metrics it is
enough to determine them for the metric
\eq
ds^2=-\sinh^2 p dt^2+ dp^2+ d\mu^2
\eqx
where we have
\eqn
\sinh p &=& \f{\cot u}{\sqrt{f(s)}} \\
d\mu &=& {\phi_{as}\over \gamma} \sqrt{f(s)} ds
\eqnx
The geodesic equations read
\eqn
t''+2\coth p\; p' t' &=& 0 \\
p''+\cosh p \sinh p\; t'^2 &=& 0 \\
\mu'' &=& 0
\eqnx
and the null condition is
\eq
\mu'^2+p'^2=\sinh^2 p\; t'^2
\eqx
It is convenient to parameterize the geodesics with $p$ instead of the affine
parameter $\lm$. One can check that the solution is
\eqn
t(p) &=& c_2+ {\rm arctanh} \f{c_1 \cosh p}{\sqrt{c_1^2-\sinh^2 p}} \\
\mu(p) &=& \arctan \f{\cosh p}{\sqrt{c_1^2-\sinh^2 p}} +c_3 \\
dp/d\lm &=& \f{\sqrt{c_1^2-\sinh^2 p}}{\sinh p}
\eqnx
As mentioned above, $p$ at the horizon is $0$, while approaching the boundary point
leads to
\eq
\sinh^2 p_A= \f{1}{\sinh^2 x}
\eqx
in agreement with
(\ref{e.bndryid}).
In order to determine which geodesic to take we have to fix the direction of the
null geodesic emitted from the boundary. In the following we utilize the $(u,\om)$ coordinates  which are defined in appendix \ref{appendc}. Near the boundary
this requirement means that we follow a curve of fixed $\om$. We have to translate
this condition to a condition on the derivative
\eq
\f{d\mu}{dp}
\eqx
at $p=p_A$. To this end we have
\eq
\f{d\mu}{dp}=\f{d\mu}{ds} \f{ds}{dp}={\phi_{as}\over \gamma}\sqrt{f(s)} \f{ds}{dp}
\eqx
where the derivatives have to be taken along constant $\om$. We can evaluate
the last derivative by differentiating
\eq
\sinh p =\f{\cot(\sqrt{\om}\sqrt{1-s})}{\sqrt{f(s)}}
\eqx
with respect to $p$ and taking the limit $s\to 1$. From this we see that $ds/dp$ is finite
which means that
\eq
\f{d\mu}{dp} \to \infty
\eqx
Evaluating $\f{d\mu}{dp}$ gives
\eq
\f{d\mu}{dp} = \f{\sinh p}{\sqrt{c_1^2-\sinh^2 p}}
\eqx
Hence this condition enables us to identify the constant
\eq
c_1=\sinh p_A
\eqx
Now the geodesic takes the following form in the $\mu$-$p$ plane:
\eq
\mu=\arctan \f{\cosh p}{\sqrt{\sinh^2 p_A-\sinh^2 p}}-\f{\pi}{2}+\mu_0
\eqx
where $\mu_0$ corresponds to $\phi=1$.
Evaluating the above expression at the horizon we get
\eq
\mu_H=\arctan \f{1}{\sinh p_A} -\f{\pi}{2}+\mu_0
=\arctan \sinh x -\f{\pi}{2}+\mu_0
\label{m2map}
\eqx
This is the boundary-horizon map based on null geodesics. By inspection
we see that it is different from the one in (\ref{mapmethod1}).

If we take $x\to \infty$ ($\om \to \infty$), and we recover $\mu_H\to \mu_0$
which is the expected result. However one may be puzzled by the opposite limit.
When $\om \to 0$ which should correspond to $x\to 0$, $\mu_H$ does not approach
0 but rather
\eq
\mu_0-\f{\pi}{2} 
\eqx
As an aside, from the general formula for $dp/d\mu$ we see that at the horizon
the geodesic is always perpendicular to the horizon so the latter condition does
not allow us to discriminate between geodesics.


Based on the above boundary-horizon map in (\ref{m2map}), the horizon length is given by
\eq
L_H =2\,\int_0^{\mu_H}d\mu\sqrt{f(\mu)}=\log
\left[\,
{1+{\rm sn}(\kappa_+ \mu_H, k^2)\over
1-{\rm sn}(\kappa_+ \mu_H, k^2)}\,
\right] 
\label{lengthm2}
\eqx
We also rewrite (\ref{m2map}) as
\eq
\sin(\mu_0-\mu_H)= 1/\cosh x_A=1/\cosh {L\over 2} 
\eqx
From this formula, we would like to identify the leading order correction
of the entropy.
Noting
\eq
\mu_0={\pi\over 2}\Big(1+{3\over 8}\gamma^2\Big)+O(\gamma^4) 
\eqx
one finds
\eq
\sin \mu_H= \tanh x_A + {3\pi\over 16}{\gamma^2\over \cosh x_A}+ \cdots
\eqx
In addition, the Jacobi sine function can be expanded as
\eq
{\rm sn}(\kappa_+ \mu_H, k^2)=
\sin \mu_H +{\gamma^2\over 8}\Big(
-3 \mu_H \cos \mu_H +\sin \mu_H \cos^2 \mu_H
\Big)+\cdots
\eqx
From this, one finds
\eq
L_H =L +{\gamma^2\over 4}\Big[
\tanh {L\over 2} + 3\cosh {L\over 2}\, \arcsin \Big({1\over \cosh {L\over 2}}\Big)
\Big]+O(\gamma^4) 
\eqx
Interestingly,
this leads to
\eq
\Delta S= {\gamma^2\over 4G}+O(\gamma^4) 
\eqx
as $L$ becomes infinity. Thus we obtained the expected result
to this order.
Let us now rewrite (\ref{lengthm2}) as
\eq
L_H =\log
\left[\,
{{\rm dn}(\kappa_+ (\mu_0-\mu_H), k^2)+{\rm cn}(\kappa_+ (\mu_0-\mu_H), k^2)\over
{\rm dn}(\kappa_+ (\mu_0-\mu_H), k^2)-{\rm cn}(\kappa_+ (\mu_0-\mu_H), k^2)}\,
\right] 
\eqx
Then using
\eq
\mu_0-\mu_H= \arcsin \Big({\rm sech} {L\over 2}\Big) 
\eqx
one finds
\eq
L_H \rightarrow L+\log {1\over \sqrt{1-2\gamma^2}} 
\eqx
Again
this leads to
\eq
\Delta S_{L=\infty} ={1\over 4G}\,\log {1\over \sqrt{1-2\gamma^2}} 
\label{ientropy}
\eqx
which
 agrees with the result of the holographic entanglement.

\subsection{Remarks on the agreement between the two methods}

Let us discuss now what aspects of the boundary-horizon map are probed
by the above calculations of the agreement between the entanglement entropy
and the two computations using the two choices of the boundary-horizon map.

If the size of the system would be finite (with a circle compactification), then the total entropy would be
obviously completely independent of the choice of the boundary-horizon map
as it has a purely geometrical definition as the area (length in the 3D case)
of the horizon. In the case of the infinite system that we consider,
we have to subtract off the extensive contribution -- and hence a
potential difference may arise only in the upper limit of integration of the
horizon area element. Therefore we are probing differences between the
boundary-horizon maps in the asymptotic near BTZ region. Unfortunately,
so far we do not have a test which would be sensitive to the finer details
of the boundary-horizon map closer to the defect.

\subsection{First law of thermodynamics}

From the previous investigation, the entropy of the system in the large size
limit is given by
\eq
S= {\pi TL\over 4 G} + S_I
\eqx
where
$S_I$ denotes $\Delta S_{L=\infty}$ in (\ref{ientropy}) 
and we have recovered the temperature dependence.
$S_I$ is temperature independent and can be interpreted as
the localized interface contribution to the entropy.
From the stress tensor in (\ref{stress}), the energy and pressure can be identified
as
\eq
E= {\pi T^2 L\over 4 G} 
\ \ \ \ \ \ p={\pi T^2 \over 4 G} 
\eqx
Therefore one can check that the first law of thermodynamics
\eq
TdS = dE +p dL
\eqx
holds precisely for our Janus system. Of course one should note that our investigation is valid
only for the large size
limit.

\section{Conclusion}\label{sec7}

In the present paper we have considered a supergravity dual to a three-dimensional
interface conformal field theory at finite temperature. The supergravity fields which
are turned on are the metric and the dilaton.
The interface is realized as the boundary between two domains ($x>0$ and $x<0$)
with differing values of the vacuum expectation value of the operator dual to the
dilaton $\pm \phi_{as}$.

The undeformed finite temperature case corresponds to the well-known BTZ black hole.
We have started from a linearized analysis of the scalar perturbation with small $\phi_{as}$
of the BTZ black hole building up intuitions concerning the general structure
of the solution including the extraction of the boundary energy-momentum tensor
which is quite intricate in coordinates natural for the Janus solution.

Using this knowledge we have formulated a scheme suitable for the numerical computation
of the exact Janus black hole for arbitrary $\phi_{as}$, the key obstacle being
an \emph{a-priori} lack of knowledge about the coordinate domain of the exact nonlinear
solution. We overcame this problem by linking a spatial coordinate to the value of
the dilaton.

In the case of three-dimensional system we found an exact analytical solution of
the finite-temperature Janus black hole. Let us emphasize that, due to the coupling
between gravity and the scalar field, the three-dimensional Einstein-dilaton
system is nontrivial in contrast to pure three-dimensional gravity.
The existence of an analytical exact solution was in fact completely unexpected
for us.

This exact solution is very interesting from various points of view.
On the general relativity side it provides an example of a black hole
in equilibrium with a nonuniform horizon. It would be very interesting to
explore such features as Hawking radiation and temperature in this setting.

From the AdS/CFT perspective, such a black-hole gives a dual description of
a three-dimensional interface CFT at finite temperature, which may have condensed matter applications.
On a more theoretical side, such a nonuniform black hole provides
a theoretical laboratory for investigating various issues dealing with entropy
of the dual field theory. In particular, we can probe various maps between
boundary points and horizon points which have been proposed in the literature,
and critically examine the problem whether local entropy density can be defined
at all.

As a step in this direction we have evaluated the total entropy deviation from
the BTZ answer using two different boundary-bulk map prescriptions finding
agreement with entanglement entropy calculations.

There are numerous directions for further research. First,
one may also
consider
ICFT defined on a circle with two interface points. Of course there is a corresponding
Janus solution dual to this  compact version of ICFT. This system can
serve as an ideal setup to study finite
size effects of a finite temperature ICFT.
 Secondly one can investigate the
gravitational aspects of the nonuniform horizon, generalize the setup to higher
number of dimensions (where probably one would have to rely on numerical methods).
One new nontrivial feature in higher  dimensions may be the appearance of
nonuniform energy density which is kinematically ruled out in the three-dimensional
case.    It would also be an interesting question to explore  if and how  the presence of an interface modifies the occurrence of a phase transition (such as a confinement/deconfinement transition) in AdS/CFT.

\section*{Acknowledgement}
We would like to thank Shiraz Minwalla, Tadashi Takayanagi and Hyunsoo Min for helpful discussions.
DB was
supported in part by NRF SRC-CQUeST-2005-0049409 and  NRF Mid-career Researcher
Program 2011-0013228. MG was supported in part by NSF grant PHY-07-57702. RJ was
supported by Polish science funds as a research project N N202 105136 (2009-2012).

\appendix
\section{Computation of entropy correction based on the conformal perturbation theory}
In this appendix we are interested in computing  free energy defined by
\eq
\beta F= -\log {\rm tr}\, e^{-\beta H} 
\eqx
Perturbing the above from $H_0$ of BTZ system, one
has
 \eq
 \beta F= \beta F_0 +\gamma^2 \beta F_{2}+O(\gamma^4) 
 \eqx
The $O(\gamma)$ contribution vanishes because
\eq
\langle {\cal L}(-i\tau,x) \rangle=0 
\eqx
where the expectation value is evaluated with respect to the BTZ system with $H_0$.
And one can find
\eq
\beta F_2 ={1\over 2} \int^\beta_0 d\tau \int^\beta_0 d\tau' \int^\infty_{-\infty}
dx \epsilon(x) \int^\infty_{-\infty}
dx' \epsilon(x')\langle {\cal L}(-i\tau,x) {\cal L}(-i\tau',x')\rangle
\eqx
where  $\epsilon(x)$ is the sign function.
Note that the two point function \cite{Bak:2007qw} is given by
\eq
\langle {\cal L}(-i\tau,x) {\cal L}(-i\tau',x')\rangle = {1\over 16 \pi^2 G}{1\over \big[\cos(\tau-\tau')-
\cosh(x-x')-i\epsilon\big]^2} 
\eqx
for the BTZ background  with $\beta=2\pi$.
Using the integral
\eq
\int_0^{2\pi} dx {1\over (\cos x +Q)^2}= 2\pi {Q\over (Q^2-1)^{3/2}} 
\eqx
$\beta F_2$ can be rearranged as
\eq
\beta F_2 =-{1\over 4 G} \int_0^{\infty}dx \int^{\infty}_{-\infty} dx' \epsilon(x')
{\cosh(x-x')+q^2\over\big((\cosh(x-x')+q^2\big)^2-1)^{3/2}}  
\eqx
with $q^2=i\epsilon$.
Using the symmetry of the integrand, the integral may be further arranged as
\eq
\beta F_2 =-{1\over 2 G} \int_0^{L/2}dx \int^{x}_{0} dx'
{\cosh x'+q^2\over\big((\cosh x'+q^2\big)^2-1)^{3/2}}  
\label{regint}
\eqx
where we also introduce a finite system size $L$. This regularized integral is finite
and has the expansion,
\eq
\beta F_2 = c_{-1}{L\over q^2} +{c_{-{1\over 2}}} {1\over q} + c_0(L) + O(q)
\eqx
where $L$ dependence in the leading term is introduced by the dimensional ground.
$c_{-1}$ is not physical since it reflects our choice of the vacuum energy level.
We evaluate the integral (\ref{regint}) by the replacement of the integrand by
\eq
\beta {\cal F}_2 =-{1\over 2 G} \int_0^{L/2}dx \int^{x}_{0} dx'
{(1+q^2)\cosh x'\over\big((1+q^2)\sinh^2x'+2q^2+q^4\big)^{3/2}}  
\eqx
which does not affect $c_{-1}$, $c_{-{1\over 2}}$ and $c_0$ terms in the expansion.
This fact may be checked numerically, whose details are in Appendix B.
Thus, by the replacement,  the physical term $c_0(LT)$ is not affected.

Then the integral may be carried out explicitly leading to
\eq
\beta {\cal F}_2 ={1\over 2G}{1+q^2\over 2q^2+q^4}\log{(q^2+1)\cosh{{L\over 2}}
+\sqrt{(q^2+1)\cosh^2{L\over 2}+q^4 +q^2-1}
\over {1+q^2}+\sqrt{2q^2+q^4}
} 
\eqx
which can be expanded as
\eq
\beta {\cal F}_2 = -{1\over 2G}\Big[
{L\over 4q^2}-{1\over \sqrt{2}q}+ {1\over 2}\coth{L\over 2}+O(q)
\Big] 
\label{icft}
\eqx
Therefore one  has
\eq
\Delta S= 
{\gamma^2\over 4G}+O(\gamma^4) 
\eqx
as the size $L$ becomes large.

Note that our computation here does not care about possible finite size
effect. Namely any effect of  boundary conditions at $x=\pm L/2$ was not incorporated
into the computation above.
Including  the finite size effects on  the field theory side is an interesting open problem.

\section{Equivalence of $F_2$ and ${\cal F}_2$ to $O(q^0)$}
We are interested in the integral
\eq
J(q, L)=
\int_0^{L/2}dx \int^{x}_{0} dx' {\cal I}(q,x')
\eqx
where
\eq
{\cal I}(q,x)=
{\cosh x+q^2\over\big((\cosh x+q^2\big)^2-1)^{3/2}} 
\eqx
We will show that
\eq
J(q, L)={L\over 4q^2}-{1\over \sqrt{2}q}+ {1\over 2}\coth{L\over 2}+O(q) 
\label{approx}
\eqx
Using an integration by parts and a change of variables, one may rearrange
$J(q, L)$  as
\eq
J(q, L)=\int^{L\over 2}_0 dx \,\Big[\,{L\over 2}-x\,\Big]\,{\cal I}(q,x)
=q\int^{L\over 2q}_0 dy \,\Big[\,{L\over 2}-q y\,\Big]\,{\cal I}
(q,q y) 
\eqx
We then expand the integrand into a power series in $q$ and perform the
integral order by
order. And then expand the result again into a power series in $q$, which
leads to the series expansion of (\ref{approx}) in $L$.
For instance by expanding the integrand to the order $q^6$, we get
\eq
J(q, L)={L\over 4 q^2}-{1\over \sqrt{2}q}+
{1\over L}+{L\over 12}-{L^3\over 720}+{L^5\over 30240}-
{L^7\over 1209600}+O(q) 
\eqx
which agrees with the $L$ expansion of (\ref{approx}).
This check can be pushed further to the higher orders in $L$.

\section{The boundary metric and coordinates of exact solution}\label{appendc}

In this appendix we shall construct the boundary metric from the exact solution (\ref{e.exact}).
For this we have to pass
to the Fefferman-Graham coordinates but just at the leading order.
In order to approach the boundary we will take the limit $u\to 0$ in a correlated
manner with $\varphi \to 1$ namely
\eq
\varphi=1-\f{1}{\om} u^2
\eqx
keeping $\om$ fixed. From the exact metric, one finds
\eq
w={ 2\phi_{as}\over \gamma \sinh^2 p } 
\label{bdid}
\eqx
The exact Janus BH metric reads then in the $(u,\om)$ coordinates at leading
order in $u$:
\eq
ds^2=\f{d\tau^2}{u^2}+\f{du^2}{u^2}+\f{H(\om) d\om^2}{u^2}
\eqx
where
\eq
H(\om)=\f{1}{4\om(\om+
{2\phi_{as}\over \gamma}
)}
\eqx
The boundary coordinate
is then found from
\eq
\sqrt{H(\om)} d\om=dx 
\eqx
Straightforward integration leads to
\eq
x=\log \Big( {\sqrt{\gamma \om}+\sqrt{\gamma\om+{2\phi_{as}} 
}
}\Big)
\eqx
which can be inverted to yield
\eq
\om= {2\phi_{as}\over \gamma}
\, \sinh^2 x  
\eqx
Finally using (\ref{bdid}), one finds
\eq
{1\over \sinh^2 p} = \sinh^2 x  
\eqx

\end{document}